\documentclass[9pt,twocolumn,twoside]{optica}

\usepackage{graphicx,amssymb,mathtools,mathptmx} 

\newcommand{\Tr}{\mathop{\mathrm{Tr}} \nolimits}
\newcommand{\re}{\mathop{\mathrm{Re}} \nolimits}
\newcommand{\Section}[1]{\par\medskip\par\noindent\textbf{#1}%
  \qquad\ignorespaces}
\makeatletter
\renewcommand{\@oddfoot}{\relax}
\renewcommand{\@evenfoot}{\relax}
\renewcommand{\@oddhead}{\hfil\thepage}
\renewcommand{\@evenhead}{\thepage\hfil}
\makeatother

\title{Quantum Fisher Information with Coherence\rule{0pt}{35pt}}

\author[1]{Zden\v{e}k~Hradil}
\author[1]{Jaroslav~\v{R}eh\'a\v{c}ek}
\author[2,3]{Luis~S\'{a}nchez-Soto} 
\author[4,5,6]{Berthold-Georg Englert}
\affil[1]{Department of Optics, Palack\'y University, 17.\ listopadu 12,
  771 46 Olomouc, Czech Republic}

\affil[2]{Departamento de \'Optica, Facultad de F\'{\i}sica, Universidad
  Complutense, 28040~Madrid, Spain}  

\affil[3]{Max-Planck-Institut f\"ur die Physik des Lichts, Staudtstra\ss e 2,
  91058 Erlangen, Germany} 

\affil[4]{Centre for Quantum Technologies, Singapore 117543, Singapore}

\affil[5]{Department of Physics, National University of Singapore, Singapore
  117551, Singapore}

\affil[6]{MajuLab, International Joint Research Unit UMI 3654, %
    CNRS, Universit{\'e} C{\^o}te d'Azur, Sorbonne Universit{\'e}, %
  National University of Singapore, Nanyang Technological University, Singapore}

\dates{Posted on the arXiv on 22 October 2019; %
   this final version on 12 November 2019}  

\doi{}%{https://doi.org/10.1364/OPTICA.6.001437}

\begin{document}%\thispagestyle{empty}

\begin{abstract}
In recent proposals for achieving optical super-resolution, variants of the
Quantum Fisher Information (QFI) quantify the attainable precision.
We find that claims about a strong enhancement of the resolution resulting
from coherence effects are questionable because they refer to very small
subsets of the data without proper normalization.
When the QFI is normalized, accounting for the strength of the signal, there
is no advantage of coherent sources over incoherent ones.
Our findings have a bearing on further studies of the achievable precision of
optical instruments. 
\end{abstract} 

\setboolean{displaycopyright}{false}
\maketitle

\Section{Introduction}
Estimation and detection theory, formulated originally as a useful tool
for signal analysis and efficient parameter estimation, became
indispensable in quantum information processing, where the effects are subtle,
signals are sparse, and  any wasting of information is detrimental. 
However, these well-established techniques can be used even in classical
detection schemes, with robust signals pushing the resolution to ultimate
limits that have not been fully explored as yet.

Recent research pioneered by Tsang and collaborators, and inspired by a
reconsideration of the classical Rayleigh criterion for the resolution of
optical instruments such as telescopes or microscopes, has received considerable
attention in the optical community (see \cite{Tsang-review} and references
therein).  
The problem can be paraphrased:
How well can we distinguish two bright spots? 
The celebrated Rayleigh arguments suggest that this can be done up to the
distances when two blurred spots start to overlap.
This rule of thumb can be justified by an analysis of the Fisher information
for the intensity pattern, and one finds that the Fisher information vanishes
for zero separation.   
As shown by Tsang and coworkers \cite{Tsang} and demonstrated experimentally
\cite{Opt}, this behavior can be avoided if quantum estimation theory is
adopted for the estimation of geometrical parameters, namely the transversal
separation and the centroid positions of two equally bright spots with known 
intensities.
In this context, the Fisher information refers to quantum measurements and
becomes the Quantum Fisher Information (QFI) upon optimizing over all
thinkable measurement schemes. 

As shown in \cite{Multiparameter}, however, the model used in \cite{Tsang} is
not robust with respect to the inclusion of other parameters.
When the intensities of the bright spots are considered as estimated
parameters, together with the separation and the centroid, the QFI remains
constant only if the two intensities are equal, but it drops to zero for
unequal intensities. 
The unphysical situation of exactly equal intensities is singular and
exhibits anomalous features. 

The ongoing research on the estimation of optical effects addresses also the
possible coherence of optical signals, and a recent discussion did not reach a
consensus \cite{Saleh,Comment_Tsang,Reply_Larson}.
Whereas the paper \cite{Saleh} claims that the presence of coherence yields a
QFI that vanishes for zero separation, the comment \cite{Comment_Tsang} shows
by explicit calculations that this need not be so.
The argument somehow paradoxically sticks to Rayleigh's reasoning for
incoherent image processing  instead of applying the Sparrow resolution limit
\cite{Sparrow} and its modifications \cite{Mehta, Cesini,Asakura}, which is
the appropriate tool for quantifying the performance of (partially) coherent
systems. 
According to the Sparrow criterion, two point sources can just be
resolved when the second derivative of the image intensity vanishes at the
point mid-way between the overlapping images of the two points.
Particularly remarkable is the argumentation in favor of using an ``anti-phase''
superposition \cite{Cesini}:
``Since the amplitude impulse response is an even function, zero intensity
results at the mid-point between the two images whatever is  the value of the
separation. This suggests that, under ideal conditions, infinite resolution is
approached.''   

In this Letter, we explain the reasons for these misunderstandings on the
basis of simple physical arguments and explicit calculations for an elementary
model of a coherent superposition.  
Our central observation is quite simple:
When coherence effects are taken into account, the Fisher information itself
is no longer a meaningful measure of accuracy, because the channels exhibiting
interference are not equivalent with respect to the strength of the signal.
Indeed, the (Quantum) Fisher Information $F$ quantifies the content of
information per registered particle; the Cr\'amer--Rao inequality (here for a
single parameter $\theta$),
\begin{equation}
  (\Delta \theta )^2\geq H \equiv \frac{1}{{nF}}\,,
\label{precision}
\end{equation}
sets a bound on the precision $H$ with which $\theta$ can be estimated from
the data.
Here, $(\Delta \theta )^2$ is the expected value of the variance of the
estimator, and $n$ is the number of detected particles.
This number is just as important as the Fisher information in the product $nF$.

We recall that the
Cr\'amer--Rao bound on the precision in (\ref{precision}) is subject to
two specific assumptions: (i) The estimator is unbiased; and (ii) the
detection events are uncorrelated, they are independent and identically
distributed (i.i.d.) random events. 
As a consequence of assumption (i) we have the unit numerator, while
assumption (ii) is crucial for the product $nF$ in the denominator---the
single-event Fisher information is multiplied by the number of detection
events.
One needs to verify that both assumptions are true in the situation of
interest. 
Further, estimation is always model-dependent and, therefore, one must check
the ingredients of the model that is used. 
We take for granted that all these verifications have been done.

\Section{Method and Results}
Let us now elaborate on the argumentation for an ideal equal-weight
superposition of symmetrically displaced sources.
We  phrase what follows in a quantum parlance, so a wave of complex amplitude
$U(x)$ can be assigned to a ket $|U\rangle$, such that
$U(x)=\langle x|U\rangle$, where $\langle{x}|$ is the bra for a point-like
source at $x$. 
The quantum formulation (using these bra and ket symbols) facilitates the
optimization since the intensity detection (and the corresponding complex
amplitudes) need not represent an optimal scheme.
More specifically, we denote by $\Psi(x)=\langle x|\Psi\rangle$ the
amplitude of the (generic) point-spread function (PSF) of the coherent
spatially-invariant imaging system.  
The coherence matrix relevant for the discussion is
\begin{eqnarray}\label{Phi-rho}
   \varrho=\frac{1}{N}|\Phi\rangle\langle\Phi|\quad\text{with}\quad
      |\Phi\rangle=|\Psi_{+}\rangle+e^{i\varphi}|\Psi_{-}\rangle\,. 
\end{eqnarray}
Here, $|\Psi_{\pm}\rangle=\exp(\pm iPs/2)|\Psi\rangle$ are the spatially
shifted PSF amplitudes, generated by the momentum operator
$P$, $\langle x|P=-i\partial_x\langle x|$,  and
$N=\langle\Phi|\Phi\rangle=2[1+\re\bigl(e^{-i\varphi}\langle
\Psi|e^{isP}|\Psi\rangle)]$ is the  normalization.
For notational simplicity, we do not indicate the dependence on the
separation $s$ and the relative phase $\varphi$ in the superposition ket
$|\Phi\rangle$, in $N$, or in~$\rho$. 

It is important to note that we are not dealing with a genuine quantum
problem.
We are using the quantum formalism for classical optics.
The PSF amplitude $\Psi(x)$ is not a probability amplitude but a classical
quantity, such as a component of the electric field.
Therefore, $\Psi(x)$ is real, and the corresponding distribution for $P$ is
even, so that $\langle f(P)\rangle=\langle\Psi|f(P)|\Psi\rangle=\langle
f(-P)\rangle$ for all functions of $P$.
In particular, then, $\langle P\rangle=0$ and there is no difference
between $\langle P^2\rangle$ and the momentum variance of the PSF,
${(\Delta P)^2=\langle P^2\rangle-\langle P\rangle^2}$.

The QFI for the parameter $s$ can be calculated from the rank-1 expression
$F=2\Tr[(\partial_{s}\rho)^2]$, which yields
\begin{equation}
  F_\varphi(s) =
  \frac{4}{N}  \langle\partial_{{s}}\Phi|\partial_{{s}}\Phi\rangle  -
  \frac{4}{N^2}\bigl|\langle\Phi|\partial_{{s}}\Phi\rangle\bigr|^2\,.
 \label{QFI}
\end{equation}
Note that it is \emph{as if} the $s$-dependence of the normalization $N$ is
ignored; in fact, its various contributions take care of each other.
The analysis of the role of coherence in the parameter estimation
hinges upon this expression for the QFI.

For understanding the role of coherence, the moment expansion for a small
displacement $s\to0$ is essential.
A complication arises since $|\Phi\rangle=0$ when both $s=0$ and
$\varphi=\pi$, so that $\rho$ is ill-defined in this limiting situation of
destructive interference at vanishing separation, and it matters whether the
limit $\varphi\to\pi$ succeeds or precedes the limit $s\to0$.
For $s=0$ and $\varphi\neq\pi$, one obtains
\begin{equation}\label{F0}
    F_{\varphi}(0)=\tan^2(\varphi/2)(\Delta P)^2\,.
\end{equation}
The QFI at $s=0$ is clearly diverging for $\varphi\to\pi$.

\begin{figure}
  \includegraphics[width=1\columnwidth]{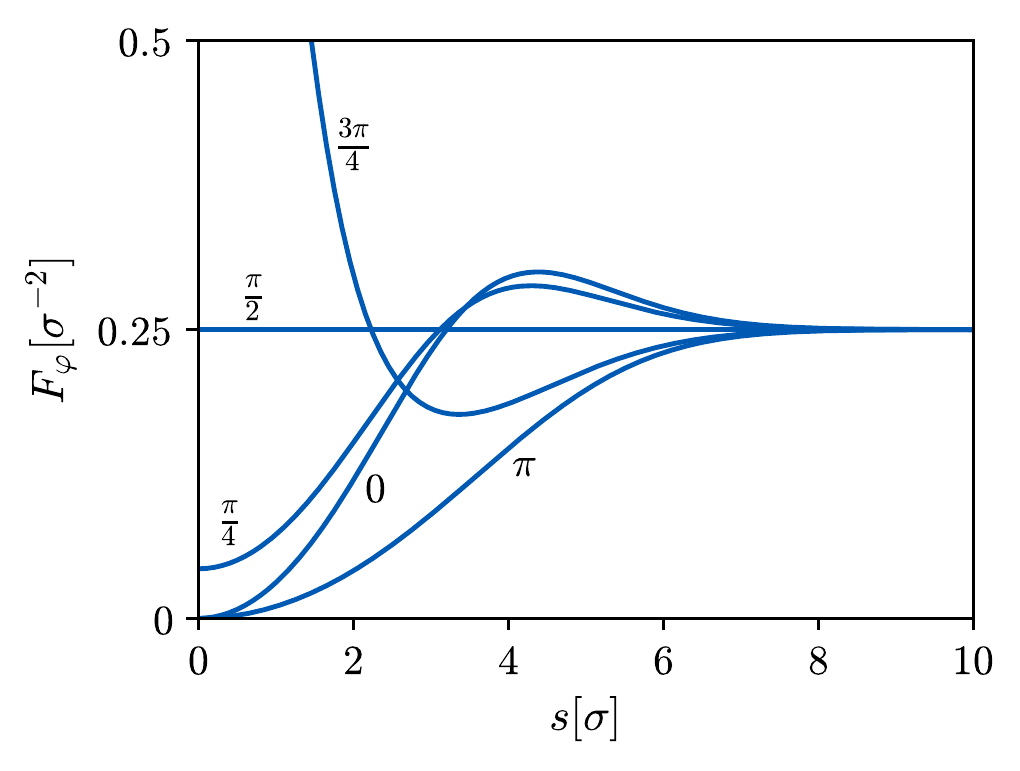}
  \caption{\label{figFish}%
    Dependence of the QFI on the displacement $s$ for the phase values
    $\varphi=0$, $\frac14 \pi$, $\frac12\pi$, $\frac34 \pi$, and $\pi$.
    The QFI diverges for $\varphi\to\pi$.
    The plot is for a Gaussian PSF with the variances
    $(\Delta X)^2=\sigma^2$  and $(\Delta P)^2=1/(4\sigma^2)$.
    The displacement is in units of $\sigma$, and $F$ in units of
    $\sigma^{-2}$.}
\end{figure}

For $\varphi=\pi$, we have
\begin{equation}\label{Fpi}
  F_{\pi}(s)=\frac{\langle P^6\rangle\langle P^2\rangle
                   -\langle P^4\rangle^2}{36\langle P^2\rangle^2}s^2    
\end{equation}
for $s$ values so small that the terms proportional to $s^4,s^6,\dots$ can be
ignored.
Also for $\varphi=0$, the leading small-$s$ contribution is not given by
(\ref{F0}); rather we have
\begin{equation}
  \label{F0s}
  F_0(s)=\frac{1}{4}\bigl(\Delta P^2\bigr)^2s^2\,,
\end{equation}
which involves the variance of $P^2$.
Figure \ref{figFish} shows $F_{\varphi}(s)$ as a function of $s$ for several
$\varphi$ values, for a Gaussian PSF.
  
It is amusing to note that the QFI for the coherent superposition
with $\varphi=\pi/2$ equals exactly the limit of incoherent
mixtures. 
This illustrates nicely Goodman's observation \cite{Goodman} that
``when (coherent) sources are in quadrature, the image intensity
  distribution is identical to that resulting from incoherent point sources.'' 

%%%%%%%%%%%%%%%%%%%%%%%%%%%%%
%%%%%%%%%%%%%%%%%%%%%%%%%%%%%

\Section{Discussion}
Understanding the behavior of the QFI for $\varphi=\pi$ is essential for the
correct interpretation of the role of coherence in estimation problems.   
The QFI clearly exhibits a singularity when $s\to0$ in this situation of
destructive interference. 
Physically speaking, we are detecting a signal on a dark fringe, where the
intensity is extremely low. 
If the norm $N=\langle\Phi|\Phi\rangle$ is taken as a weighting factor into
the definition of the precision $H$ in (\ref{precision}), the singularity
disappears.  
There are plausible and sound physical arguments for the inclusion of such a
weight: 
constructive and destructive coherence is always manifested by an enhancement
or a suppression of the emerging signal and this represents a valuable resource,
which should be taken into account.
This argument can be supported by an exact calculation of the cost of
preparing the superposition in (\ref{Phi-rho}).   
The analysis can be linked to state-of the art technology \cite{nature} for
the deterministic generation of these superpositions.

\begin{figure}
  \includegraphics[width=1\columnwidth]{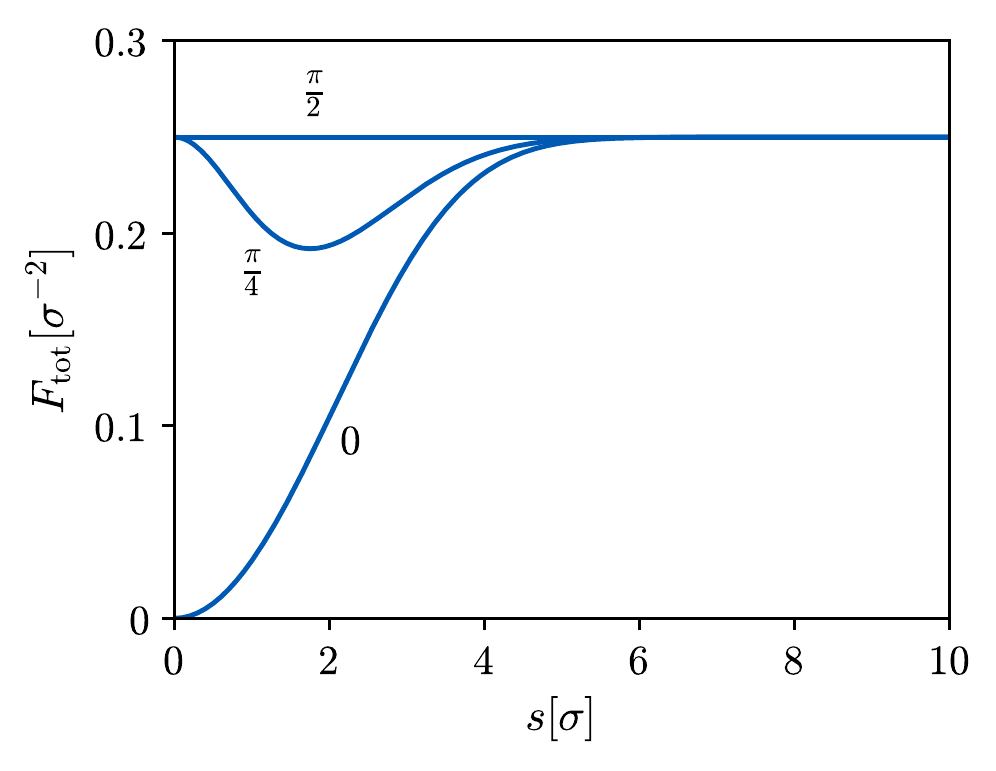}
  \caption{\label{figMZ}%
    Dependence of the total QFI on the displacement $s$ for both constructive
    and destructive interference channels detected independently for phases
    $\varphi =0$, $\frac14 \pi$, and $\frac12 \pi$.
    This properly normalized  QFI  is always limited by its
    value for an incoherent superposition.
    Note that the upper bound is saturated for large displacements,
    and also for zero displacement for all the phases
    except when $\varphi=0$ or $\pi$.
    The plot is for a Gaussian PSF  with the variances
    $(\Delta x)=\sigma^2$ and $(\Delta P)^2=1/(4\sigma^2)$.}
\end{figure}

We generate the superposition in (\ref{Phi-rho}) from an entangled
state,
\begin{eqnarray}\label{entangled}
  |\varphi\rangle&=&2^{-1/2}\bigl(|\Psi_+\rangle
                     \otimes|\mathnormal{\uparrow}_x\rangle
    +e^{i\varphi}|\Psi_-\rangle\otimes|\mathnormal{\downarrow}_x\rangle\bigr)
  \nonumber\\&=&|\Phi_1\rangle\otimes|\mathnormal{\uparrow}_z\rangle
               +|\Phi_2\rangle\otimes|\mathnormal{\downarrow}_z\rangle\,,
\end{eqnarray}
where $|\Phi_1\rangle$ is half of the superposition in (\ref{Phi-rho}) and
$|\Phi_2\rangle$ is that for $\varphi\to\varphi+\pi$,
\begin{equation}
  \label{Phi12}
  |\Phi_1\rangle=\frac{1}{2}\bigl(|\Psi_{+}\rangle
                                  +e^{i\varphi}|\Psi_{-}\rangle\bigr)\,,
  \quad
  |\Phi_2\rangle=\frac{1}{2}\bigl(|\Psi_{+}\rangle
                                   -e^{i\varphi}|\Psi_{-}\rangle\bigr)\,,
\end{equation}
and $|\mathnormal{\uparrow}_{x,z}\rangle$,
$|\mathnormal{\downarrow}_{x,z}\rangle$ are the (pseudo-)spin states of an
auxiliary qubit.
The entanglement here is that available in classical light \cite{Qian-Eberly}.
The desired  coherent superpositions are obtained upon measuring the qubit in
the $|\mathnormal{\uparrow}_{z}\rangle,|\mathnormal{\downarrow}_{z}\rangle$
basis, and the probabilities of occurrence are given by the norms of the
respective superpositions, $\langle\Phi_1|\Phi_1\rangle$ and
$\langle\Phi_2|\Phi_2\rangle$.
If the probability that matters is very small, as is the case when
dark-fringe data are selected, the procedure has a large overhead of
discarded data, and a fair assessment cannot ignore these costs.
Accordingly,  there are three ways how to asses the
QFI for such a generic scheme: 
\begin{itemize}
\item[(E)]
   From a joint measurement on the system and the qubit, for which the
   QFI is $F_{\text{ent}}=\langle P^2\rangle=(\Delta P)^2$,
   obtained by applying (\ref{QFI}) to the entangled state $|\varphi\rangle$
   in (\ref{entangled}).
   While the optimal measurement may not be feasible, as it will require the
   distinction of entangled states, the value of $F_{\text{ent}}$ is an
   upper bound on the QFI from any other procedure.
 \item[(I)]
   From the entangled state $|\varphi\rangle$, with the qubit traced
   out---only the system is measured. 
   The resulting QFI is that for the incoherent mixture,
   $F^{\ }_{\text{inc}}=(\Delta P)^2$.
   Since $F^{\ }_{\text{inc}}=F_{\text{ent}}$, the entangled-basis
   measurements of scheme (E) offers no actual advantage.
 \item[(S)]
   From measurements that are conditioned on finding the qubit in the state
   $|\mathnormal{\uparrow}_{z}\rangle$ or in the state
   $|\mathnormal{\downarrow}_{z}\rangle$.
   In this case, the data are sorted into two sub-ensembles, and their QFIs
   have to be weighted by their respective probabilities of occurrence to
   yield the total QFI, 
\begin{eqnarray}
  F^{\ }_{\text{tot}}(s)&=&
  \langle\Phi_1|\Phi_1\rangle F_{\varphi}(s)
          +\langle\Phi_2|\Phi_2\rangle F_{\varphi+\pi}(s)\nonumber\\
  &\leq& \langle P^2\rangle=F^{\ }_{\text{ent}}=F^{\ }_{\text{inc}}\,.
    \label{Fsorted}
\end{eqnarray}
  If the data of one sub-ensemble are ignored, then the respective term is
  removed from the sum, while the remaining term continues to be weighted by
  its probability of occurrence.
  As an immediate consequence of (\ref{F0}),
  together with $\langle\Phi_1|\Phi_1\rangle\bigr|_{s\to0}=\cos^2(\varphi/2)$
  and $\langle\Phi_2|\Phi_2\rangle\bigr|_{s\to0}=\sin^2(\varphi/2)$,
  we have
  $F^{\ }_{\text{tot}}(0)=(\Delta P)^2=F^{\ }_{\text{inc}}$ except for
  $\phi=0$ or $\pi$.
  Figure \ref{figMZ} shows $F^{\ }_{\text{tot}}(s)$ as a function of $s$ for
  several $\varphi$ values, for a Gaussian PSF.
\end{itemize}
In view of this argumentation, it is clear that the coherence, although the
QFI may diverge for one sub-ensemble, does not provide any improvement over an
incoherent source, if the cost of generating such a signal is properly
taken into account.
Nothing is gained by an increase of the factor $F$ in the product $nF$ in
(\ref{precision}) if the value of $n$ decreases even more. 

%%%%%%%%%%%%%%%%%%%%%%%%%%%%%
%%%%%%%%%%%%%%%%%%%%%%%%%%%%%

Other sorting schemes than scheme (S) can also be realized with the option of
having situations intermediate between the fully coherent and the completely
mixed sub-ensembles. 
In the case of partial coherence, the explicit form of the partially coherent
state matters for the QFI of this sub-ensemble.
The properly weighted total QFI for \emph{any} sorting scheme, however, cannot
exceed the upper bound set by $F_{\text{ent}}=F_{\text{inc}}$.
This observation supports the arguments used in \cite{Saleh} and, so we think,
settles the discussion in \cite{Saleh,Comment_Tsang,Reply_Larson}.

For simplicity, the discussion above deals with the estimation of a single
parameter, the separation $s$, which is sufficient for demonstrating the case,
namely that the sub-ensembles carry weights and these weights enter the total
QFI in (\ref{Fsorted}).
When the data from an actual experiment are evaluated, however, the
multi-parameter situation of asymmetrically displaced sources, with
unequal intensity and partial coherence, matters.
Then, the sub-ensembles are not specified by pure states like those in
(\ref{Phi12}), but by rank-2 states of the generic form
\begin{equation}
  \label{rank-2}
  \rho_j=\bigl(\begin{array}{cc}|\Psi_+\rangle & |\Psi_-\rangle\end{array}\bigr)
R_j \Biggl(\begin{array}{c}\langle\Psi_+| \\ \langle\Psi_-|\end{array}\Biggr)\,,
\end{equation}
where now $|\Psi_{\pm}\rangle=\exp\bigl(iP(s_0\pm s/2)\bigr)|\Psi\rangle$
and $R_j$ is a hermitian $2\times2$ matrix restricted by $\rho_j\geq0$ and
$\mathrm{tr}(\rho_j)=1$.
In addition to the separation $s$ and the location parameter $s_0$, there are
further parameters specifying the $R_j$s in accordance with the model considered.
For each parameter, we have the QFI $F_j$ of the $j$th sub-ensemble, and the
properly weighted sum of the $F_j$s replaces $F^{\ }_{\text{tot}}$ of
(\ref{Fsorted}).
Together with the corresponding value of the count $n$, this yields the analog
of the product $nF$ in the Cr\'amer--Rao bound (1), provided that the usual
conditions are met; in particular, the estimators must not be biased.
The situations discussed in \cite{Saleh,Comment_Tsang,Reply_Larson} are
particular cases of this multi-parameter scenario. 

Finally, concerning the  so-called ``Rayleigh curse'' --- a term
coined in \cite{Tsang}, where the value of the QFI at vanishing separation is
by itself regarded as a significant measure for distinguishability,
and $F(0)=0$ is the poor ``classical'' resolution
(yes curse) while $F(0)>0$ is the superior ``quantum'' resolution (no curse)
--- we observe a few points.
First, the estimator for the displacement $s$ usually exhibits a substantial
bias when $s$ is small, and then the Cr\'amer--Rao bound of (\ref{precision})
does not apply without the necessary modification. 
Second, the product $nF$ is relevant in (\ref{precision}), not just the Fisher
information, and nothing is gained by an increase of $F$ if it is compensated
for by a decrease of $n$; while an individual QFI in the sum in
(\ref{Fsorted}) can easily exceed $F_{\text{inc}}$, the properly weighted sum
cannot. 
 
\Section{Concluding remarks}
The simple model studied here is sufficient to make the point that the QFI is
but one ingredient and that there is no genuine advantage of coherent over
incoherent sources when all aspects are accounted for. 
The model is good enough to explain the discrepancies in the analysis of
coherent effects in \cite{Saleh,Comment_Tsang,Reply_Larson}.
But the model has its obvious limitations in that only one parameter is
considered (the separation), and any analysis of a realistic situation has
to deal with at least two more parameters, namely the centroid position and
the relative intensity of the two sources.
While there could be more parameters of relevance, such as the degree of
coherence, certainly these three need to be estimated jointly from the data.
A realistic analysis must also pay close attention on how the parameters are
estimated from the data; the biases and the mean-square errors (or any
other measure of accuracy) of the estimators actually used
matter in practice, not the Cr\'amer--Rao bound for optimal unbiased
estimators.
Clearly, much more work is needed before the community can reach a definite
conclusion about the benefits of coherent sources, or coherent procedures
for data acquisition, for the resolution of optical instruments.

\Section{Dedication}
We dedicate this paper to the memory of Helmut Rauch (1939--2019), the pioneer
of neutron interferometers, who taught us so much about the importance of
coherence.

\section*{Funding Information}
ZH and J\v{R} acknowledge financial support from the Grant Agency of the
Czech Republic (Grant No. 18-04291S),
LSS acknowledges financial support from the Spanish MINECO
(Grant FIS2015-67963-P).
The Centre for Quantum Technologies is a Research Centre of Excellence funded
by the Ministry of Education and the National Research Foundation of
Singapore.

\newcommand{\DOI}[2]{\href{#1}{#2}}


\begin{thebibliography}{14}
 
\bibitem{Tsang-review}
M. Tsang,
``Resolving starlight: a quantum perspective,''
\DOI{https://arxiv.org/abs/1906.02064}%
{e-print arXiv:1906.02064[quant-ph] (2019).}
 
\bibitem{Tsang}
M. Tsang, R. Nair, and X.-M. Lu,
``Quantum theory of superresolution for two incoherent optical point sources,''
\DOI{https://doi.org/10.1103/PhysRevX.6.031033}%
{\JournalTitle{Physical Review X}\ \textbf{6}, 031033 (2016).}

\bibitem{Opt}
M. Pa\'ur, B. Stoklasa, Z. Hradil, L. L. S\'anchez-Soto,
and J. \v{R}eh\'a\v{c}ek,
``Achieving the ultimate optical resolution,''
\DOI{http://doi.org/10.1364/OPTICA.3.001144}%
{\JournalTitle{Optica}\ \textbf{3}, 1144--1147 (2016).}

\bibitem{Multiparameter}
J. \v{R}eh\'a\v{c}ek, Z. Hradil, B. Stoklasa, M. Pa\'ur, J. Grover, A. Krzic,
and L. L. S\'anchez-Soto,
``Multiparameter quantum metrology of incoherent point sources: Towards
  realistic superresolution,''
\DOI{http://doi.org/10.1103/PhysRevA.96.062107}%
{\JournalTitle{Physical Review A}\ \textbf{96}, 062107 (2017).}

\bibitem{Saleh}
W. Larson and B. E. A. Saleh,
``Resurgence of Rayleighs curse in the presence of partial coherence,''
\DOI{http://doi.org/10.1364/OPTICA.5.001382}%
{\JournalTitle{Optica}\ \textbf{5}, 1382--1389 (2018).}

\bibitem{Comment_Tsang}
M. Tsang and R. Nair,
``Resurgence of Rayleigh's curse in the presence of partial coherence:
  comment,''
\DOI{https://doi.org/10.1364/OPTICA.6.000400}%
{\JournalTitle{Optica}\ \textbf{6}, 400--401 (2019).}

\bibitem{Reply_Larson}
W. Larson and B. E. A. Saleh,
``Resurgence of Rayleigh's curse in the presence of partial coherence:
  reply,''
\DOI{https://doi.org/10.1364/OPTICA.6.000402}%
{\JournalTitle{Optica}\ \textbf{6}, 402--403 (2019).}

\bibitem{Sparrow}
C. M. Sparrow,
``On spectroscopic resolving power,''
\DOI{https://doi.org/10.1086/142271}%
{\JournalTitle{Astrophysical Journal}\ \textbf{44}, 76--86 (1916).}

\bibitem{Mehta} 
B. L. Mehta,
``Two point resolution with non-uniform and non-symmetric
  illumination using partially coherent light,''
\DOI{https://doi.org/10.1088/0335-7368/5/2/304}%
{\JournalTitle{Nouvelle Revue d'Optique}\ \textbf{5}, 95--99 (1974).}

\bibitem{Cesini} 
G. Cesini, G. Guattari, P. De Santis, and C. Palma,
``Two-point resolution with anti-phase coherent illumination.
  I. One-dimensional systems,''
\DOI{https://doi.org/10.1088/0150-536x/10/2/004}%
{\JournalTitle{Journal of Optics}\ \textbf{10}, 79--87 (1979).}

\bibitem{Asakura}
T. Asakura,
``Resolution of two unequally bright points with partially coherent light,''
\DOI{https://doi.org/10.1088/0335-7368/5/3/304}%
{\JournalTitle{Nouvelle Revue d'Optique}\ \textbf{5}, 169--177 (1974).}

\bibitem{Goodman}
J. W. Goodman,
\textit{Introduction to Fourier Optics}
(Roberts and Company Publishers, 3rd edition, 2005), p.~159.

\bibitem{nature} 
B. Hacker, S. Welte, S. Daiss,  A. Shaukat, S. Ritter, L. Li, and G. Rempe,
``Deterministic creation of entangled atom-light Schr\"odinger cat states,''
\DOI{https://doi.org/10.1038/s41566-018-0339-5}%
{\JournalTitle{Nature Photonics}\ \textbf{13}, 110--115 (2019).}

\bibitem{Qian-Eberly}
X.-F. Qian and J. H. Eberly,
``Entanglement and classical polarization states,''
\DOI{https://doi.org/10.1364/OL.36.004110}%
{\JournalTitle{Optics Letters}\ \textbf{36}, 4110--4112 (2011).}

\end{thebibliography}
\end{document}